\documentclass[10pt]{article}
\usepackage{fancyhdr}
\usepackage{extramarks}
\usepackage{amsmath}
\usepackage{amsthm}
\usepackage{amsfonts}
\usepackage{siunitx}
\usepackage{tikz}
\usepackage[plain]{algorithm}
\usepackage{algpseudocode}
\usepackage{multirow}
\usepackage{booktabs}
\usepackage{graphicx}
\usepackage{subfigure}
\usepackage[colorlinks,linkcolor=black,anchorcolor=black,citecolor=black,urlcolor=blue]{hyperref}
\usepackage{amsmath,bm}
\usepackage{booktabs}
\usepackage{mathtools}
\usepackage{amssymb}
\usepackage{caption}
\usepackage{capt-of}
\usepackage{makecell}
\usepackage{mciteplus}
\usepackage{cite}
\usepackage{mathrsfs}
\usepackage[title,titletoc,toc]{appendix}
\usepackage{xr}
\usepackage{parskip}
\usepackage{soul}
\usepackage{textcomp}
\usepackage[colaction]{multicol}
\usepackage[switch]{lineno}
\usepackage{lipsum}
\usepackage{etoolbox}
\usepackage{longtable}
\usepackage{array}
\usepackage{tablefootnote}
\usepackage{ragged2e}
\newcolumntype{C}[1]{>{\centering\arraybackslash}p{#1}}
\usetikzlibrary{automata,positioning}
\usetikzlibrary{automata,positioning}

\makeatletter
\newcommand*{\addFileDependency}[1]{
	\typeout{(#1)}
	\@addtofilelist{#1}
	\IfFileExists{#1}{}{\typeout{No file #1.}}
}
\makeatother

\begin{document}

\title{Machine-learning Repurposing of DrugBank Compounds for Opioid Use Disorder  
}

\author{ Hongsong Feng$^1$, Jian Jiang$^{2,1}$, and Guo-Wei Wei$^{1,3,4}$\footnote{
		Corresponding author.		Email: weig@msu.edu} \\
	$^1$ Department of Mathematics, \\
	Michigan State University, MI 48824, USA.\\
	$^2$ Research Center of Nonlinear Science, School of Mathematical and Physical Sciences,\\
	Wuhan Textile University, Wuhan, 430200, P R. China\\
	$^3$Department of Electrical and Computer Engineering,\\
	Michigan State University, MI 48824, USA. \\
	$^4$ Department of Biochemistry and Molecular Biology,\\
	Michigan State University, MI 48824, USA.  
}

\date{\today} 

\maketitle

Opioid use disorder (OUD) is a chronic and relapsing condition that involves the continued and compulsive use of opioids despite harmful consequences. The development of medications with improved efficacy and safety profiles for OUD treatment is urgently needed. Drug repurposing is a promising option for drug discovery due to its reduced cost and expedited approval procedures. Computational approaches based on machine learning enable the rapid screening of DrugBank compounds, identifying those with the potential to be repurposed for OUD treatment. We collected inhibitor data for four major opioid receptors and used advanced machine learning predictors of binding affinity that fuse the gradient boosting decision tree algorithm with two natural language processing (NLP)-based molecular fingerprints and one traditional 2D fingerprint. Using these predictors, we systematically analyzed the binding affinities of DrugBank compounds on four opioid receptors. Based on our machine learning predictions, we were able to discriminate DrugBank compounds with various binding affinity thresholds and selectivities for different receptors. The prediction results were further analyzed for ADMET (absorption, distribution, metabolism,  excretion, and toxicity), which provided guidance on repurposing DrugBank compounds for the inhibition of selected opioid receptors. The pharmacological effects of these compounds for OUD treatment need to be tested in further experimental studies and clinical trials. Our machine learning studies provide a valuable platform for drug discovery in the context of OUD treatment.


\pagenumbering{roman}
\begin{verbatim}
\end{verbatim}

\newpage
\clearpage
\pagebreak
{\setcounter{tocdepth}{4} \tableofcontents}
\newpage

\setcounter{page}{1}
\renewcommand{\thepage}{{\arabic{page}}}

		\section{Introduction}

		
		Opioid use disorder (OUD) is a chronic illness characterized by periods of relapse and remission, causing significant distress and impairment \cite{mclellan2000drug}. Despite the harmful consequences, individuals with opioid addiction experience strong cravings to obtain and use opioids. In the United States, three million people have experienced or are currently struggling with OUD, while globally the number is 16 million \cite{dydyk2022opioid}. The opioid epidemic has raised public awareness and led to significant resources being invested in combating the problem. Effective medications and behavioral interventions have played a key role in preventing relapse, promoting longer periods of abstinence, and reducing mortality and morbidity \cite{douaihy2013medications}.

		
		Currently, the US Food and Drug Administration (FDA) has approved three medications for treating Opioid Use Disorder (OUD): methadone, buprenorphine, and naltrexone. These medications have been shown to be effective in treating addiction by acting on different opioid receptors, each with varying pharmacological effects. The three major targets for opioids are the mu (MOR), kappa (KOR), and delta (DOR) opioid receptors. MOR is associated with the euphoria and rewarding properties of stimuli, as well as the maintenance of drug use, drug craving, and relapse \cite{wang2019historical}. KOR, on the other hand, can produce dysphoric effects and trigger anti-reward effects \cite{wee2010role}. This receptor has the opposite effect of MOR, and KOR antagonists may be beneficial in treating depressive disorders, especially among individuals with addiction \cite{knoll2010dynorphin}. DOR can induce anxiolytic effects and attenuate depressive symptoms \cite{roberts2001increased}. Two other important receptors, the nociceptin opioid receptor (NOR) and opioid growth factor receptor (ZOR), have not been studied extensively.
		
		Methadone is a full agonist of MOR and has therapeutic effects in reducing withdrawal and craving symptoms. Due to its longer half-life and fewer drug-like effects, such as euphoria, it causes fewer withdrawal symptoms and is less reinforcing \cite{brown2004methadone}. Methadone can decrease the intensity of craving, which is beneficial for patient adherence to treatment. However, methadone is associated with the risk of respiratory depression in overdose. Buprenorphine is a partial MOR agonist and KOR antagonist. Both methadone and buprenorphine can suppress opioid withdrawal, attenuating the effects of injected opioids, and protecting against overdose \cite{bell2014pharmacological}. To mitigate the risk of misuse, it is often prescribed in combination with naloxone, an opioid antagonist that can block its effects. As a partial agonist, buprenorphine has a ceiling effect of euphoria and reduced risks of respiratory depression compared to methadone \cite{mattick2014buprenorphine}. Naltrexone, a MOR and KOR antagonist, has limited utility for OUD treatment due to poor adherence \cite{morgan2018injectable}. However, naltrexone's KOR antagonist properties could contribute to mood improvements in OUD patients \cite{weerts2008differences}. Long-acting injectable naltrexone has been reported to reduce cravings. Naloxone is an opioid antagonist used to reverse respiratory depression in opioid overdose. Due to its high affinity, naloxone can displace opioid drugs such as heroin, fentanyl, or morphine, interfering with their respiratory depressant effects. Naloxone and naltrexone are antagonists that target all opioid receptor subtypes. Unlike agonists, they do not stimulate the opioid receptor and therefore have no pharmacological effects such as sedation, analgesia, respiratory depression, and euphoria \cite{mclellan2000drug}.

		
		Although current medications are effective in treating OUD, relapse and remission remain common due to neurobiological changes and opioid receptor tolerance caused by repeated opioid abuse \cite{wang2019historical}. The traditional process of developing new drugs is expensive and time-consuming, taking over a decade and costing billions of dollars. However, advances in technology and innovation in drug development, such as drug repurposing, can accelerate this process. Drug repurposing, which investigates an existing drug for a new therapeutic indication other than its original intended purpose, has been successful in several instances and can reduce development costs and timelines \cite{patwardhan2016innovative}. To identify repurposable drug candidates, computational and experimental approaches have been proposed and used synergistically. Computational approaches, such as genetic association, molecular docking, pathway mapping, signature matching, and retrospective clinical analysis, are cost-effective and promote the discovery of more repurposable candidates \cite{pushpakom2019drug,hegde2015unravelling, singh2017drug}. One promising computational method for drug development, including drug repurposing, is machine learning, which can learn from available biochemical data and predict drugs' clinical indications, toxicity profiles, or therapeutic potentials \cite{kim2019drug, napolitano2013drug, gilvary2020machine}. With the accumulation of genetic and structural databases, machine learning algorithms can accelerate the discovery of new repurposable drugs.

		
		Machine learning predictions offer a promising approach for investigating drug repurposing as a potential strategy for developing effective medications for OUD treatment. As previously mentioned, opioid receptors, specifically MOR, KOR, and DOR, are crucial targets for developing medications to treat OUD. Binding affinity is a crucial factor for the efficacy of drugs in exerting their pharmacological effects. Several approved agonist/antagonist medications for OUD treatment have demonstrated high affinities for displacing opioids from the receptors, which can alleviate withdrawal symptoms and drug cravings \cite{blanco2019management}.
		
		Machine learning techniques can be utilized to predict the binding effects of compounds to different opioid receptors and identify potential repurposable drugs. The ChEMBL database \cite{gaulton2017chembl} contains inhibitor compounds with experimentally determined binding labels that serve as the foundation for machine learning predictions of binding affinity. The development of molecular descriptors has been enhanced by natural language processing (NLP) \cite{chen2021extracting, winter2019learning} and mathematically-based 3D representations \cite{feng2023virtual}. These approaches can extract important molecular physical and stereochemical information, thus improving the descriptive and predictive ability of machine learning methods for virtual screening of small molecules and facilitating drug repurposing predictions.

		
		DrugBank (version 5.1.10) is a publicly available database of pharmacological agents that has a collection of 8865 compounds, including 1806 approved drugs and 7059 investigational or off-market drugs \cite{wishart2018drugbank}. It provides convenient access to comprehensive molecular information about current drugs and their mechanisms, as well as  interactions with targets, which facilitates efficient drug discovery and development. It has been extensively used in repurposing studies to find therapeutic candidates for various diseases \cite{lesmana2022genomic,singh2017drug,adikusuma2021drug}, including COVID-19 \cite{mahdian2020drug,gao2020repositioning}, with in silico or in vitro investigations. A recent study proposed a machine-learning modeling approach to identify new analgesic opioids with the aid of the PubChem portal and DrugBank \cite{jia2021construction}.


		In this work, to assess the potential of repurposing DrugBank compounds as medications for OUD, we collected four sets of inhibitor data for the four main opioid receptors: MOR, KOR, DOR, and NOR. Using this curated data, we constructed machine learning (ML) models with two NLP-based fingerprints generated by transformer and autoencoder models, as well as one traditional 2D fingerprint ECFP. These models exhibited strong predictive power in the five-fold cross-validation tests. With these ML models, we systematically examined the binding affinities of these DrugBank drugs with various binding thresholds. As the pharmacological effects of drugs in treating OUD are closely linked to the specific functions of different opioid receptors \cite{wang2019historical}, we conducted systematic ML predictions and analysis to identify the DrugBank compounds that can be selectively active at specific opioid receptors. Based on these predictions, we identified a few drugs with satisfactory binding energies at certain binding affinity thresholds or selectivities. We also performed molecular docking for a few promising drugs to understand their interactions with target opioid receptors. Furthermore, we employed ML-based models to evaluate some of the FDA-approved OUD drugs or our ML-predicted promising drugs on their pharmacokinetic properties, including absorption, distribution, metabolism, excretion, and toxicity (ADMET). With satisfactory inhibition affinities on the receptors and desired ADMET properties, these identified drugs require further animal experiments to evaluate their safety and efficacy in treating OUD. Our machine learning studies yield valuable results for drug discovery in OUD treatment.

		\section{Results}
		
		\begin{figure}[ht]
			\centering
			\includegraphics[width=0.75\linewidth]{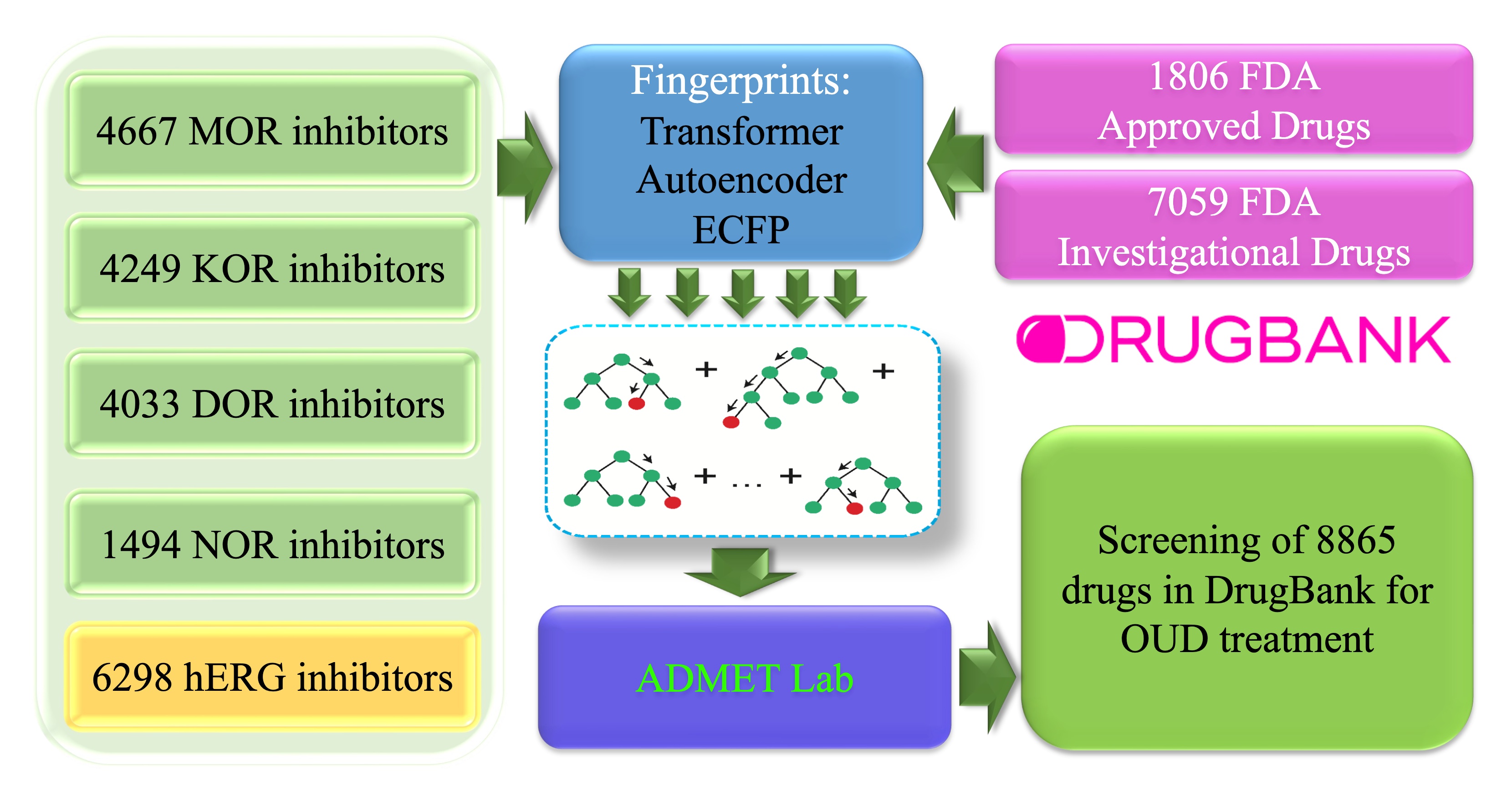} 
			\caption{{\footnotesize Schematic illustration of our machine learning platform for screening the DrugBank database for the treatment of opioid use disorder. Inhibitor datasets of MOR, KOR, DOR, NOR, and hERG were used to building machine-learning (ML) predictors of binding affinity. Three molecular fingerprints from transformer, autoencoder, and ECFP were paired with gradient boosting decision algorithm to build the ML predictors. A total of 8865 DrugBank drugs were screened for their potency on the major opioid receptors, hERG side effects and ADMET properties.
			} }
			\label{Fig:work-flow}
		\end{figure} 
		
		Figure \ref{Fig:work-flow} outlines the workflow of our machine learning-based virtual screening of the DrugBank database for the treatment of opioid use disorder. Our study builds  four  machine learning models for opioid receptors MOR, KOR, DOR, and NOR. These models are used to screen 8865 compounds in the DrugBank database for their potency. The resulting promising compounds are further screened for hERG side effects and ADMET pharmacokinetic  properties.

		\subsection{Opioid receptors and binding affinity predictors}

		MOR, KOR, DOR, and NOR are four crucial opioid receptors that have a significant impact on the development of OUD. These receptors are the primary pharmacological targets for the three FDA-approved drugs for OUD treatment: methadone, buprenorphine, and naltrexone. These medications are categorized as opioid replacement therapy (ORT) drugs, which involve substituting an opioid with a less euphoric opioid that has a longer duration of action. Naloxone is another medication commonly used to mitigate the risk of overdose.

		Our aim was to identify drugs with potential for use in opioid replacement therapy (ORT) from the DrugBank \cite{wishart2018drugbank} database. To achieve this, we collected inhibitor datasets for the four major opioid receptors from the ChEMBL database and built machine learning (ML) models to predict the binding effects of DrugBank compounds on these receptors. Additionally, we considered the hERG side effect in our analysis, as it is a critical potassium channel that must be avoided during drug discovery due to the risk of fatal arrhythmia \cite{sanguinetti2006herg}. To build an appropriate binding affinity (BA) predictor for hERG, we collected a hERG inhibitor dataset. Our machine learning models demonstrated reliable predictive power, achieving Pearson correlation coefficients (R) of 0.842, 0.862, 0.866, and 0.783, and root mean square errors (RMSE) of 0.969, 0.960, 0.944, and 0.961 kcal/mol for the MOR, KOR, DOR, and NOR inhibitor datasets, respectively. The hERG dataset model showed an R of 0.786 and an RMSE of 0.773 kcal/mol in the five-fold cross validation. Using these models, we systematically evaluated the binding affinities of approved and investigational or off-market drugs in the DrugBank database while avoiding drugs that may have side effects on hERG.

		In this study, our focus was on identifying DrugBank compounds that could potentially act as inhibitors of MOR, KOR, or DOR, since these targets are primarily modulated by approved or commonly used medications. To ensure safety, we used a binding affinity (BA) threshold of -8.18 kcal/mol ($K_i=1\mu$M) to screen out compounds with potential hERG side effects. For identifying potent inhibitors of opioid receptors, we applied thresholds of -9.54 kcal/mol ($K_i=0.1\mu$M), which is a widely accepted threshold for high affinity\cite{flower2002drug}, as well as more stringent thresholds of -10 kcal/mol and -11 kcal/mol.

		\subsection{Potential inhibitors of opioid receptors in the DrugBank}

		In order to find potential drugs for OUD treatment, we utilized our machine learning predictors to evaluate the binding affinities of drugs in the DrugBank database on the three opioid receptors. Since the 3D structures of these receptors are highly similar, a single compound could bind to multiple opioid receptors. We also noted that FDA-approved medications for OUD treatment could affect more than one of MOR, KOR, and DOR. Our machine learning predictions identified DrugBank compounds that bound to the three receptors. The reliability scores of our machine learning predictions are included in parentheses, with higher scores indicating more reliable predictions.
		
		\subsubsection{FDA-approved drugs predicted to be effective on opioid receptors}

		Table \ref{tab:drugbank-approved} showcases 15 FDA-approved compounds in DrugBank, which our machine learning models predicted to be effective against all three opioid receptors. We distinguished the three sets of drugs using BA thresholds of -11, -10, and -9.54 kcal/mol. Here's a brief overview of the 15 drugs:

		The first four drugs listed in Table \ref{tab:drugbank-approved} are predicted to have high potency on all three receptors with binding energies less than -11 kcal/mol. Our predictions assigned them reliability scores of up to 1 on most of the opioid receptors, indicating that these drugs are included in the training sets of our machine learning models. The predicted high potency of these drugs on the three receptors suggests that they are validated opioids with proven efficacy. Buprenorphine, one of the three FDA-approved medications for medication-assisted treatment (MAT) for opioid use disorder \cite{bell2020medication}, is a partial agonist with high affinity for MOR and acts as a partial agonist or functional antagonist of KOR, as well as a weak DOR antagonist \cite{wang2019historical}. Compared to the FDA-approved drug methadone, buprenorphine has a lower ceiling effect when activating MOR, resulting in less respiratory depression \cite{wang2019historical}. It is often combined with naloxone to minimize the risk of misuse \cite{koehl2019medications}.
		
		Naldemedine is an FDA-approved drug used for treating opioid-induced constipation in adults \cite{hu2018naldemedine}. It functions as a peripherally acting antagonist of MOR, which means it blocks opioid receptors outside of the brain and spinal cord. However, its ability to penetrate the blood-brain barrier and enter the central nervous system is limited, which may render it ineffective for treating opioid dependence or addiction.
		
		Samidorphan is an opioid antagonist that is used to treat schizophrenia and bipolar disorder \cite{chaudhary2019review}. It acts as a MOR antagonist and a partial agonist at KOR and DOR \cite{bidlack2018vitro}. Research on samidorphan has also explored its potential for treating alcoholism and cocaine addiction. When combined with buprenorphine, it produces antidepressant effects without the typical MOR-related euphoria or substance dependence \cite{dhir2017investigational}.

		Levorphanol is an opioid medication that has been used to treat moderate to severe pain \cite{elks2014dictionary}. It primarily acts as an agonist of MOR but is also an agonist of DOR, KOR, and NOR \cite{gudin2016levorphanol}.
		
		Drugs five to nine in Table \ref{tab:drugbank-approved} were predicted to have binding affinities less than -10 kcal/mol but greater than -11 kcal/mol on the three receptors. Levallorphan is an antagonist of MOR and an agonist of KOR \cite{goodman1955pharmacological}. It blocks the effects of stronger agents such as morphine while simultaneously producing analgesia \cite{codd1995serotonin}, and was used to reverse the respiratory depression induced by opioid analgesics. Butorphanol is a synthetic opioid analgesic that acts as a partial agonist and antagonist at the MOR, as well as a partial agonist at the KOR \cite{gharagozlou2006pharmacological}. It exhibited effectiveness in reducing post-operative shivering and is commonly used in the management of migraines \cite{gear1999kappa}. Naltrexone is another FDA-approved drug used for the treatment of OUD. It works as a non-selective opioid receptor antagonist and renders subsequent opioid ingestion ineffective, reducing opioid withdrawal symptoms in dependent individuals \cite{wang2019historical}. Its effectiveness in the management of alcohol dependence or OUD relies on reduced cravings and feelings of euphoria associated with substance use disorder. Nalmefene is an agonist of MOR and DOR and a partial agonist of KOR. Nalmefene showed effectiveness in counteracting the respiratory depression produced by an opioid overdose \cite{park2019prevention} and is also used in the management of alcohol dependence. Hydromorphone is an opioid used to treat moderate to severe pain, with one of its major hazards being dose-related respiratory depression. Injectable hydromorphone was also found to be effective for patients with severe opioid use disorder and might provide greater benefit than the FDA-approved drug methadone \cite{bansback2018cost}. These drugs are opioids, according to their reliability scores, which are mostly up to 1 on the three receptors.
		
		Drugs ten to fifteen in Table \ref{tab:drugbank-approved} were predicted to have binding energies ranging from -10 kcal/mol to -9.54 kcal/mol. Naloxone is a non-selective and competitive opioid receptor antagonist commonly used to counteract acute opioid intoxication-induced respiratory depression \cite{park2019prevention}. Oxymorphone is an opioid analgesic indicated for the management of severe pain, including post-surgical pain. Both drugs were included in our training sets, as indicated by their reliability scores of 1. The remaining four drugs, somatostatin, pentetreotide, lurbinectedin, and dotatate, were not present in our training set. However, their predicted high binding energies suggest that they could have potential in the treatment of OUD.


		\begin{table}
			\centering
			\small
			\begin{tabular}{c| c c c c c }		
				\toprule
				&\makecell[c]{Drugbank ID} & \makecell[c]{Generic Name} &\makecell[c]{BA-MOR}	& \makecell[c]{BA-KOR} &\makecell[c]{BA-DOR} \\
				\hline 	
				1&DB00921& Buprenorphine& -12.13 (1.00)& -13.70 (1.00)& -11.93 (1.00)\\
				2&DB11691& Naldemedine& -12.05 (1.00)& -11.07 (0.69)& -12.19 (1.00)\\
				3&DB12543& Samidorphan& -13.65 (1.00)& -12.86 (1.00)& -11.57 (1.00)\\
				4&DB00854& Levorphanol& -12.98 (1.00)& -11.58 (1.00)& -11.21 (1.00)\\
				\hline 	
				5&DB00504& Levallorphan& -12.13 (1.00)& -10.91 (0.85)& -11.06 (0.89)\\
				6&DB00611& Butorphanol& -12.49 (1.00)& -13.11 (1.00)& -10.70 (1.00) \\
				7&DB00704& Naltrexone& -12.34 (1.00)& -11.81 (1.00)& -10.51 (1.00)  \\
				8&DB06230& Nalmefene& -12.40 (1.00)& -11.97 (1.00)& -10.82 (1.00)   \\
				9&DB00327& Hydromorphone& -12.82 (1.00)& -11.36 (1.00)& -10.03 (1.00)\\
				\hline 					
				10&DB01183& Naloxone& -11.53 (1.00)& -10.61 (1.00)& -9.90 (1.00)\\
				11&DB01192& Oxymorphone& -11.83 (1.00)& -10.23 (1.00)& -9.92 (1.00)\\
				12&DB09099& Somatostatin& -9.97 (0.70)& -9.64 (0.68)& -9.67 (0.70)\\
				13&DB12602& Pentetreotide& -9.81 (0.66)& -9.58 (0.64)& -9.61 (0.66)\\
				14&DB12674& Lurbinectedin& -10.19 (0.63)& -9.55 (0.62)& -9.74 (0.62)\\
				15&DB14554& Dotatate& -10.12 (0.72)& -9.55 (0.70)& -9.79 (0.71)\\				
				\bottomrule
			\end{tabular}
			\caption{Summary of FDA-approved drugs with potential to inhibit MOR, KOR, and DOR, including predicted binding affinities (measured in kcal/mol) by our ML-models, along with the reliability scores of the predictions (given in parentheses). The first three drugs correspond to predicted BAs of less than -11 kcal/mol, the first eleven correspond to BAs less than -10 kcal/mol, and all fifteen drugs have BAs less than -9.54 kcal/mol.} 
			\label{tab:drugbank-approved}
		\end{table}

		\begin{figure}[ht]
			\centering
			\includegraphics[width=0.95\linewidth]{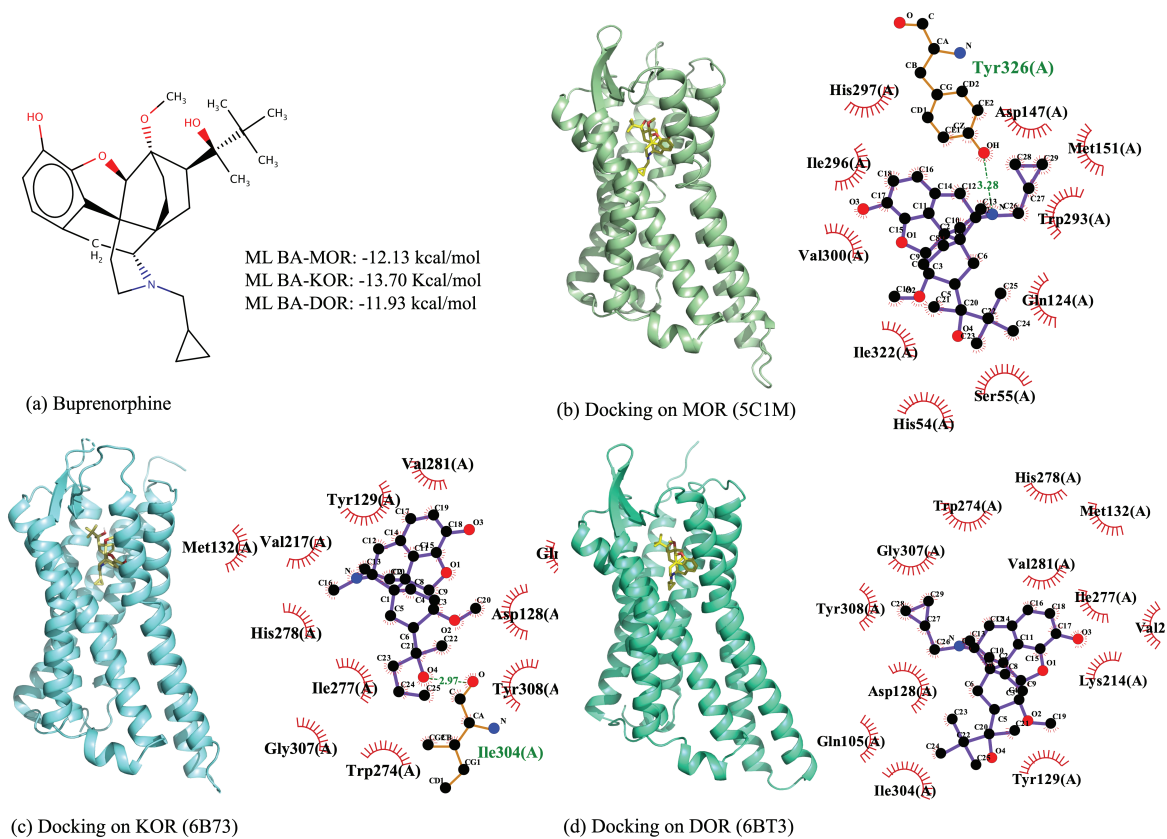} 
			\caption{{\footnotesize The docking structures of buprenorphine on the mu (MOR), kappa (KOR), and delta (DOR) receptors, along with their corresponding 2D interaction diagrams, are presented. The PDB IDs we used in the ducking analysis for the three receptors are 5C1M, 6B73, and 6BT3. } }
			\label{Fig:Docking-Buprenorphine}
		\end{figure}

		\begin{figure}[ht]
			\centering
			\includegraphics[width=0.95\linewidth]{Docking-Nalmefene.pdf} 
			\caption{{\footnotesize The docking structures of nalmefene on the mu (MOR), kappa (KOR), and delta (DOR) receptors, along with their corresponding 2D interaction diagrams, are presented. The PDB IDs we used in the ducking analysis for the three receptors are 5C1M, 6B73, and 6BT3.	
			} }
			\label{Fig:Docking-Nalmefene}
		\end{figure} 
		
		\subsubsection{Investigational drugs predicted to be effective on opioid receptors}
		
		Through our ML models, we have identified 19 investigational drugs with potential as effective inhibitors of three opioid receptors, presented in Table \ref{tab:drugbank-investigational}. These drugs were categorized based on their predicted binding affinities, with thresholds of -11, -10, and -9.54 kcal/mol. Although many of these drugs were not present in our training sets due to their reliability scores not being equal to 1, the first seven drugs have been reported to have potential uses as opioid analgesics or in the treatment of OUD. Despite the exclusion from the training sets, our models were still able to accurately distinguish these effective inhibitors, highlighting the predictive power of our ML models. Other drugs with high prediction reliability scores may also have potential as opioid medications.

		Although binding affinity is an essential factor in drug development, other properties such as toxicity, partition coefficient ($\rm \log P$), solubility ($\rm \log S$), synthesizability, pharmacodynamics, and pharmacokinetics must also be considered. Developing medications with the desired efficacy for treating OUD is a complex task, and investigational drugs may face difficulties in meeting these critical characteristics. One significant issue is the limited ability of opioid compounds to cross the blood-brain barrier and reach the opioid receptors in the central nervous system, which may limit their effectiveness in treating OUD. Additionally, many opioid medications produce euphoric effects due to their stimulation of MOR, which can prevent them from being suitable analgesics or medications for OUD. Therefore, the agonist or antagonist effects on specific opioid receptors are crucial factors in inducing therapeutic benefits of medications for OUD treatment, and further research is needed in OUD drug discovery. The FDA-approved drugs can provide better starting points for repurposing drugs, and investigational drugs require further scrutiny of their druggability profiles before being considered for clinical trials or approval.
		
		\begin{table}[ht]
			\centering
			\small
			\begin{tabular}{c| c c c c c}		
				\toprule
				&\makecell[c]{Drugbank ID} & \makecell[c]{Generic Name} &\makecell[c]{BA-MOR}	& \makecell[c]{BA-KOR} &\makecell[c]{BA-DOR} \\
				\hline 	
				
				1&DB01450& Dihydroetorphine& -11.90 (0.85)& -11.97 (0.91)& -11.74 (0.91)\\
				2&DB01480& Cyprenorphine& -12.04 (0.93)& -12.72 (0.93)& -12.04 (0.93)   \\
				3&DB01497& Etorphine& -11.36 (0.79)& -11.22 (0.87)& -11.69 (0.88)       \\
				4&DB01548& Diprenorphine& -13.00 (1.00)& -13.18 (1.00)& -12.81 (1.00)   \\
				
				\hline 	
				
				5&DB01469& Acetorphine& -10.89 (0.79)& -10.98 (0.88)& -10.83 (0.88)\\
				6&DB14682& Dextrorphan& -11.98 (0.98)& -10.57 (0.98)& -10.16 (0.98)\\
				7&DB16117& Buprenorphine hemiadipate& -10.92 (0.83)& -10.99 (0.83)& -10.64 (0.83)\\
				8&DB16072& ORP-101& -10.91 (0.80)& -10.98 (0.80)& -10.24 (0.80)\\
				9&DB04894& Vapreotide& -10.48 (0.80)& -10.08 (0.73)& -10.09 (0.75)\\
				\hline 		
				10&DB01531& Desomorphine& -11.49 (0.84)& -9.91 (0.84)& -9.73 (0.84)\\
				
				11&DB06409& Morphine glucuronide& -10.40 (1.00)& -9.66 (0.76)& -9.57 (0.78)\\
				12&DB12088& TT-232& -10.87 (0.83)& -9.69 (0.77)& -10.03 (0.78)\\
				13&DB12454& Zalypsis& -10.23 (0.61)& -9.67 (0.61)& -9.93 (0.61)\\
				14&DB15341& Dinalbuphine sebacate& -10.91 (0.86)& -10.87 (0.86)& -9.77 (0.86)\\
				15&DB15646& Fasitibant& -9.64 (0.51)& -9.57 (0.51)& -9.76 (0.51)\\
				16&DB16323& Satoreotide tetraxetan& -9.84 (0.65)& -10.01 (0.64)& -9.81 (0.69)\\
				17&DB17150& Ukrain cation& -9.88 (0.55)& -10.04 (0.54)& -9.59 (0.55)\\
				18&DB17158& Satoreotide trizoxetan& -9.65 (0.65)& -9.88 (0.64)& -9.84 (0.69)\\
				19&DB17160& Edotreotide yttrium Y-90& -10.00 (0.65)& -9.59 (0.66)& -9.60 (0.67)\\
				20&DB01721& \makecell[c]{ N-[2-hydroxy-1-indanyl]-5\\-[(2-tertiarybutylaminocarbonyl)-4(\\benzo[1,3]dioxol-5-ylmethyl)-piperazino]\\-4-hydroxy-2-(1-phenylethyl)\\-pentanamide}& -10.14 (0.62)& -9.79 (0.61)& -10.00 (0.66)\\
				\bottomrule
			\end{tabular}
			\caption{Summary of investigational or off-market drugs that have the potential to inhibit MOR, KOR, and DOR. The predicted binding affinity values (unit: kcal/mol) generated by our ML-models are provided, along with the corresponding reliability scores in parentheses. The first four, first eight, and all seventeen drugs correspond to predicted BAs of $<$ -11 kcal/mol, $<$ -10 kcal/mol, and$ <$ -9.54 kcal/mol, respectively.} 
			\label{tab:drugbank-investigational}
		\end{table}

		\begin{figure}[ht]
			\centering
			\includegraphics[width=0.95\linewidth]{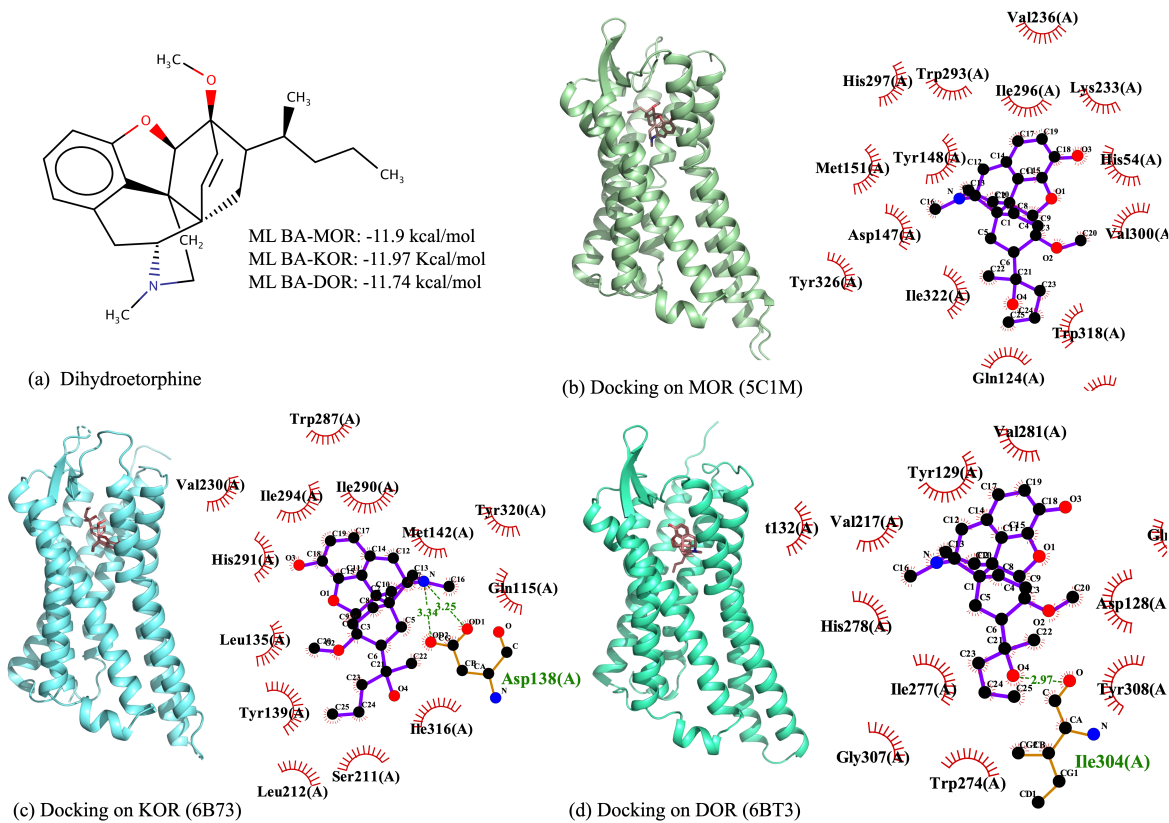} 
			\caption{{\footnotesize The docking structures of dihydroetorphine on MOR, KOR, and DOR, as well as the corresponding 2D interaction diagrams. The PDB IDs we used in the ducking analysis for the three receptors are 5C1M, 6B73, and 6BT3.} }
			\label{Fig:Docking-Dihydroetorphine}
		\end{figure} 
		
		\begin{figure}[ht]
			\centering
			\includegraphics[width=0.95\linewidth]{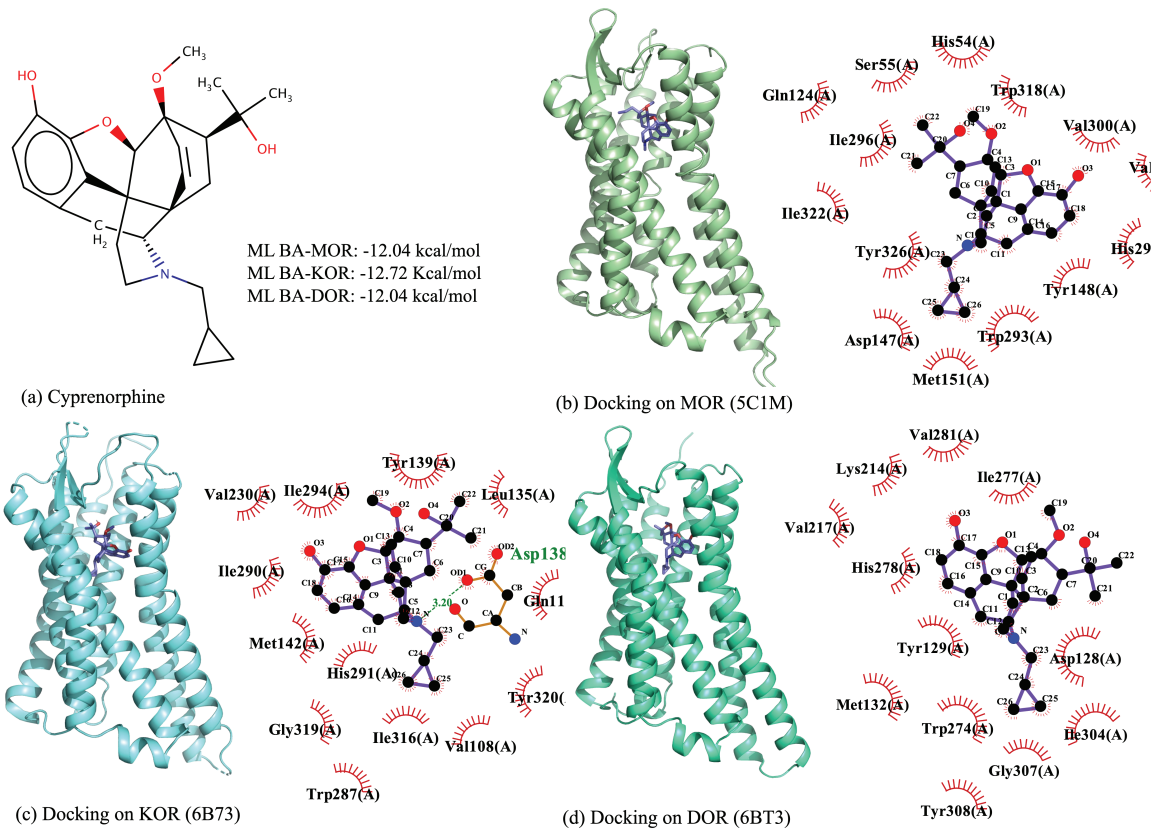} 
			\caption{{\footnotesize The docking structures of cyprenorphine on MOR, KOR, and DOR, as well as the corresponding 2D interaction diagrams. The PDB IDs we used in the ducking analysis for the three receptors are 5C1M, 6B73, and 6BT3.} }
			\label{Fig:Docking-Cyprenorphine}
		\end{figure}

		\subsubsection{Screening DrugBank  compounds with various selectivity options}
		
		The opioid receptors share high similarities in protein sequences or 3D structures, meaning that an inhibitor compound can have binding effects on multiple receptors. Table \ref{tab:drugbank-approved} and \ref{tab:drugbank-investigational} present the DrugBank compounds that were predicted to be potent at MOR, KOR, and DOR. When designing drugs, selectivity is a crucial property that enables the drug to produce the desired therapeutic effect while minimizing undesirable side effects.
		
		Using our models, we have identified several DrugBank compounds that have demonstrated selective effectiveness on one or more of the opioid receptors, namely MOR, KOR, and DOR. In addition, we have found three collections of FDA-approved drugs in DrugBank that were predicted to be potent on two of the three receptors, but weak on the remaining one. Tables \ref{tab:drugbank-approved-MOR-DOR}, \ref{tab:drugbank-approved-MOR-KOR}, and \ref{tab:drugbank-approved-DOR-KOR} present the FDA-approved drugs that were predicted to be potent at the three pairs of receptors, MOR-DOR, MOR-KOR, and DOR-KOR, respectively. A potency threshold of -9.54 kcal/mol was utilized.
		
		Table \ref{tab:drugbank-approved-MOR-DOR} includes a few drugs, such as alvimopan and eluxadoline, which were potent at MOR and DOR with similarity scores greater than 0.8 according to our predictions. These drugs have been reported to interact with opioid receptors at the molecular level, affecting activities in the nervous system. Other drugs, including morphine, pentazocine, nalbuphine, dezocine, and methylnaltrexone, listed in Table \ref{tab:drugbank-approved-MOR-KOR}, as well as tramadol in Table \ref{tab:drugbank-approved-DOR-KOR}, also showed high potency on the two receptors in each pair.
		
		In addition, some drugs, such as naloxegol, used for treating opioid-induced constipation, and gonadotropin-releasing hormone agonist drugs, including sincalide, nafarelin, triptorelin, and goserelin, showed relatively high prediction scores. However, it remains uncertain whether these hormone agonists can impact opioid receptors and function as medications for treating OUD. Tables S4 and S10 in the Supporting Information provide additional DrugBank compounds that may be selectively effective on these receptors.


		\begin{table}[ht]
			\centering
			\small
			\begin{tabular}{c| c c c c c}		
				\toprule
				&\makecell[c]{Drugbank ID} & \makecell[c]{Generic Name} &\makecell[c]{BA-MOR}	& \makecell[c]{BA-KOR} &\makecell[c]{BA-DOR}\\
				\hline 	
				1& DB00309& Vindesine& -9.72 (0.62)& -9.36 (0.65)& -9.54 (0.65)\\
				2& DB00541& Vincristine& -9.77 (0.63)& -9.30 (0.64)& -9.57 (0.65)\\
				3& DB00570& Vinblastine& -9.87 (0.63)& -9.41 (0.64)& -9.54 (0.65)\\
				4& DB04911& Oritavancin& -9.91 (0.53)& -9.36 (0.54)& -9.54 (0.54)\\
				5& DB06274& Alvimopan& -12.31 (1.00)& -9.38 (1.00)& -10.87 (1.00)\\
				6& DB06791& Lanreotide& -10.04 (0.69)& -9.51 (0.67)& -9.72 (0.67)\\
				7& DB08890& Linaclotide& -9.82 (0.61)& -9.40 (0.61)& -9.65 (0.66)\\
				8& DB09097& Quinagolide& -9.56 (0.54)& -9.51 (0.53)& -9.69 (0.54)\\
				9& DB09142& Sincalide& -9.61 (0.78)& -9.13 (0.72)& -9.84 (0.79)\\
				10& DB09272& Eluxadoline& -11.04 (0.96)& -8.85 (0.57)& -10.04 (0.58)\\
				11& DB13925& Dotatate gallium Ga-68& -10.03 (0.68)& -9.48 (0.66)& -9.77 (0.69)\\
				12& DB13985& Lutetium Lu 177 dotatate& -10.05 (0.68)& -9.48 (0.66)& -9.73 (0.68)\\
				13& DB15494& Edotreotide gallium Ga-68& -10.00 (0.64)& -9.53 (0.64)& -9.58 (0.66)\\
				14& DB15873& Copper oxodotreotide Cu-64& -9.96 (0.69)& -9.52 (0.67)& -9.71 (0.69)\\
				\bottomrule
			\end{tabular}
			\caption{Summary of the FDA-approved drugs that are potential potent inhibitors of MOR and DOR with BA $<$-9.54 kcal/mol whist being potential weak inhibitors of MOR and DOR with BA $>$-9.54 kcal/mol. The predicted binding affinities (unit: kcal/mol) by our ML-models are provided. The reliability scores of the ML-prediction are given in the parenthesis.} 
			\label{tab:drugbank-approved-MOR-DOR}
		\end{table}

		\begin{table}[ht]
			\centering
			\small
			\begin{tabular}{c| c c c c c}		
				\toprule
				&\makecell[c]{Drugbank ID} & \makecell[c]{Generic Name} &\makecell[c]{BA-MOR}	& \makecell[c]{BA-KOR} &\makecell[c]{BA-DOR}\\
				\hline 	
				1& DB00014& Goserelin& -9.58 (0.77)& -11.01 (0.82)& -8.89 (0.77)\\
				2& DB00035& Desmopressin& -10.15 (0.70)& -9.92 (0.70)& -9.31 (0.72)\\
				3& DB00199& Erythromycin& -9.65 (0.58)& -9.55 (0.60)& -8.75 (0.61)\\
				4& DB00207& Azithromycin& -9.71 (0.57)& -9.56 (0.58)& -8.93 (0.60)\\
				5& DB00295& Morphine& -11.64 (1.00)& -9.55 (1.00)& -9.12 (1.00)\\
				6& DB00520& Caspofungin& -9.83 (0.59)& -9.64 (0.59)& -8.91 (0.60)\\
				7& DB00615& Rifabutin& -9.74 (0.57)& -9.78 (0.58)& -9.13 (0.58)\\
				8& DB00644& Gonadorelin& -9.57 (0.71)& -11.02 (0.82)& -9.00 (0.71)\\
				9& DB00652& Pentazocine& -10.27 (0.83)& -10.13 (0.83)& -8.91 (0.83)\\
				10& DB00666& Nafarelin& -9.75 (0.72)& -11.04 (0.84)& -9.34 (0.74)\\
				11& DB00803& Colistin& -9.58 (0.61)& -9.72 (0.61)& -8.96 (0.63)\\
				12& DB00844& Nalbuphine& -11.23 (1.00)& -10.94 (1.00)& -9.07 (1.00)\\
				13& DB01201& Rifapentine& -9.74 (0.56)& -9.71 (0.57)& -9.43 (0.57)\\
				14& DB01209& Dezocine& -10.65 (0.81)& -9.77 (0.77)& -8.71 (0.77)\\
				15& DB01211& Clarithromycin& -9.60 (0.57)& -9.57 (0.59)& -8.80 (0.61)\\
				16& DB01301& Rolitetracycline& -10.10 (0.55)& -9.60 (0.55)& -9.49 (0.55)\\
				17& DB06663& Pasireotide& -9.75 (0.73)& -9.58 (0.73)& -9.31 (0.73)\\
				18& DB06800& Methylnaltrexone& -11.07 (0.98)& -10.47 (0.94)& -8.89 (0.98)\\
				19& DB06825& Triptorelin& -9.67 (0.74)& -10.98 (0.86)& -9.03 (0.75)\\
				20& DB09049& Naloxegol& -9.99 (0.78)& -9.80 (0.78)& -9.09 (0.78)\\
				21& DB11700& Setmelanotide& -9.98 (0.70)& -10.01 (0.69)& -9.45 (0.73)\\
				22& DB12825& Lefamulin& -9.62 (0.57)& -9.64 (0.59)& -8.77 (0.57)\\
				23& DB00781& Polymyxin B& -9.77 (0.64)& -9.73 (0.64)& -9.09 (0.65)\\
				\bottomrule
			\end{tabular}
			\caption{Summary of the FDA-approved drugs that are potential potent inhibitors of MOR and KOR with BA $<$-9.54 kcal/mol whist being potential weak inhibitors of MOR and DOR with BA $>$-9.54 kcal/mol. The predicted binding affinities (unit: kcal/mol) by our ML-models are provided. The reliability scores of the ML-prediction are given in the parenthesis.} 
			\label{tab:drugbank-approved-MOR-KOR}
		\end{table}

		\begin{table}[ht]
			\centering
			\small
			\begin{tabular}{c| c c c c c}		
				\toprule
				&\makecell[c]{Drugbank ID} & \makecell[c]{Generic Name} &\makecell[c]{BA-MOR}	& \makecell[c]{BA-KOR} &\makecell[c]{BA-DOR}\\
				\hline 	
				1& DB00193& Tramadol& -7.72 (1.00)& -10.40 (1.00)& -10.62 (1.00)\\
				\bottomrule
			\end{tabular}
			\caption{Summary of the FDA-approved drugs that are potential potent inhibitors of KOR and DOR with BA $<$-9.54 kcal/mol whist being potential weak inhibitors of MOR and DOR with BA $>$-9.54 kcal/mol. The predicted binding affinities (unit: kcal/mol) by our ML-models are provided. The reliability scores of the ML-prediction are given in the parenthesis.} 
			\label{tab:drugbank-approved-DOR-KOR}
		\end{table}
		
		Selective inhibition of KOR is a highly anticipated pharmacotherapeutic intervention for substance use disorders \cite{farahbakhsh2023systemic}. KOR has anti-reward effects in addiction, and heightened stress during withdrawal and abstinence states can enhance KOR function, leading to dysphoric mood and potential relapse \cite{wang2019historical}. Therefore, KOR antagonists could reverse dysphoria and reduce drug-seeking behavior during withdrawal and abstinence. Both naltrexone and naloxone are KOR antagonists that are effective for OUD treatment. Using our machine-learning predictions, we identified seven FDA-approved drugs and thirty-one investigational drugs that could act as KOR blockers, as shown in Tables S7 and S8.
		
		The design of KOR antagonists provides a promising direction for OUD treatment. Other therapies using opioid agonists or partial agonists may still be useful for OUD treatment. Opioid agonist therapy utilizing drugs such as methadone can reduce euphoria and withdrawal symptoms, while partial agonists like buprenorphine have a weaker effect on respiratory depression \cite{wang2019historical}. The antagonist naltrexone or naloxone is also used to improve treatment with buprenorphine. We obtained repurposing predictions of DrugBank compounds for MOR and included the promising drugs in Table S6 in the Supporting Information. In addition, DrugBank compounds predicted to be selectively potent at KOR are presented in Table s9 and S10.

		\subsubsection{Molecular interactions between opioid receptors and potent inhibitors}

		Understanding how drugs bind to their targets is crucial for comprehending the molecular mechanism of drug-target interactions \cite{vo2021mu}. To predict the binding of several FDA-approved and investigational drugs to opioid receptors, we utilized the molecular docking software AutoDock Vina \cite{trott2010autodock} in this study.
		
		Buprenorphine is an FDA-approved drug used to treat OUD that was shown to be potent at MOR, KOR, and DOR in our predictions. Figure \ref{Fig:Docking-Buprenorphine}b reveals a hydrogen bond between buprenorphine and MOR, formed by the nitrogen atom of buprenorphine and the hydrogen in the hydroxyl group on the residue Tyr326. Similarly, Figure \ref{Fig:Docking-Buprenorphine}c shows one hydrogen bond between buprenorphine and KOR, formed by the hydrogen in the hydroxyl group of the drug and an oxygen atom of residue Ile304. No hydrogen bonds were observed in the molecular interactions of buprenorphine with DOR, as shown in Figure \ref{Fig:Docking-Buprenorphine}d. Hydrophobic interactions may account for the drug's promising binding affinity with DOR.
		
		Nalmefene, an MOR antagonist and KOR partial agonist, has been demonstrated to effectively counteract opioid overdose-induced respiratory depression \cite{park2019prevention}. The docking poses displayed in Figure \ref{Fig:Docking-Nalmefene} reveal that nalmefene forms hydrogen bonds with KOR and DOR. As seen in Figure \ref{Fig:Docking-Nalmefene}c, the drug forms three hydrogen bonds with KOR, one between a hydrogen atom from a hydroxyl in the residue Tyr139 and one oxygen atom on the drug, and two between a hydrogen atom in one hydroxyl with the nitrogen atom on residual Gln115 and oxygen atom on residual Asp138 of KOR. Figure \ref{Fig:Docking-Nalmefene}d shows one hydrogen bond between a hydrogen atom in one hydroxyl on the drug and an oxygen atom on residual Asp128 in DOR. The molecular interactions between the drug and MOR are mainly hydrophobic bonds, as shown in Figure \ref{Fig:Docking-Nalmefene}a.
		
		Dihydroetorphine is a promising investigational drug that has shown potential to be effective on MOR, KOR, and DOR. This potent opioid analgesic is mainly used in China and has gained attention as a substitute maintenance drug for treating opioid addiction, similar to buprenorphine, which is widely used in western nations \cite{husbands2003opioid}. Its predicted binding affinities on MOR, KOR, and DOR are -11.9, -11.97, -11.74 kcal/mol, respectively. The docking mode of dihydroetorphine on MOR is shown in Figure \ref{Fig:Docking-Dihydroetorphine}b. Although no hydrogen bonds were found, the binding energies may be attributed to hydrophobic interactions. The docking poses of dihydroetorphine on KOR shown in Figure \ref{Fig:Docking-Dihydroetorphine}c indicate the formation of two hydrogen bonds between the N atom of the drug and hydrogen atoms from two hydroxyls on the residue Asp138 of KOR. One hydrogen bond between dihydroetorphine and DOR is shown in Figure \ref{Fig:Docking-Dihydroetorphine}d, which is formed between an oxygen atom on the drug and a hydrogen atom in a hydroxyl on the residue Ile304.
		
		Another investigational drug that has shown high potency on all three receptors is cyprenorphine, which has mixed agonist-antagonist effects on opioid receptors similar to those of buprenorphine. Studies have reported that cyprenorphine can block the binding of morphine and etorphine to these opioid receptors \cite{keep1971etorphine}. The binding modes of cyprenorphine on the three receptors are illustrated in Figure \ref{Fig:Docking-Cyprenorphine}. No hydrogen bonds were observed between the drug and two of the receptors, namely MOR and DOR, as shown in Figure \ref{Fig:Docking-Cyprenorphine}b and \ref{Fig:Docking-Cyprenorphine}d. Their binding affinities may be mainly attributed to hydrophobic interactions. However, in Figure \ref{Fig:Docking-Cyprenorphine}c, a single hydrogen bond was observed between the drug and KOR, formed by the nitrogen atom in the drug with a hydrogen atom in the hydroxyl of residue Asp138.

		\subsection{ADMET analysis}
		
		ADMET (absorption, distribution, metabolism, excretion, and toxicity) plays a critical role in drug discovery and development as it encompasses a wide range of attributes related to the pharmacokinetic studies of a compound. A drug candidate with satisfactory therapeutic efficacy must also meet appropriate ADMET criteria. Accurate predictions of ADMET could reduce the risk of late-stage failure in drug design. Therefore, to identify promising candidates for treating OUD, screening for ADMET properties is necessary. We used ADMETlab 2.0 solvers \cite{xiong2021admetlab,lei2016admet} for machine-learning predictions, and their documentation provides optimal ranges for these properties, which are detailed in Table S3 in the supporting information. We mainly focused on thirteen indexes of ADMET properties in their predictions.
		
\begin{figure}[ht]
			\centering
			\includegraphics[width=0.99\linewidth]{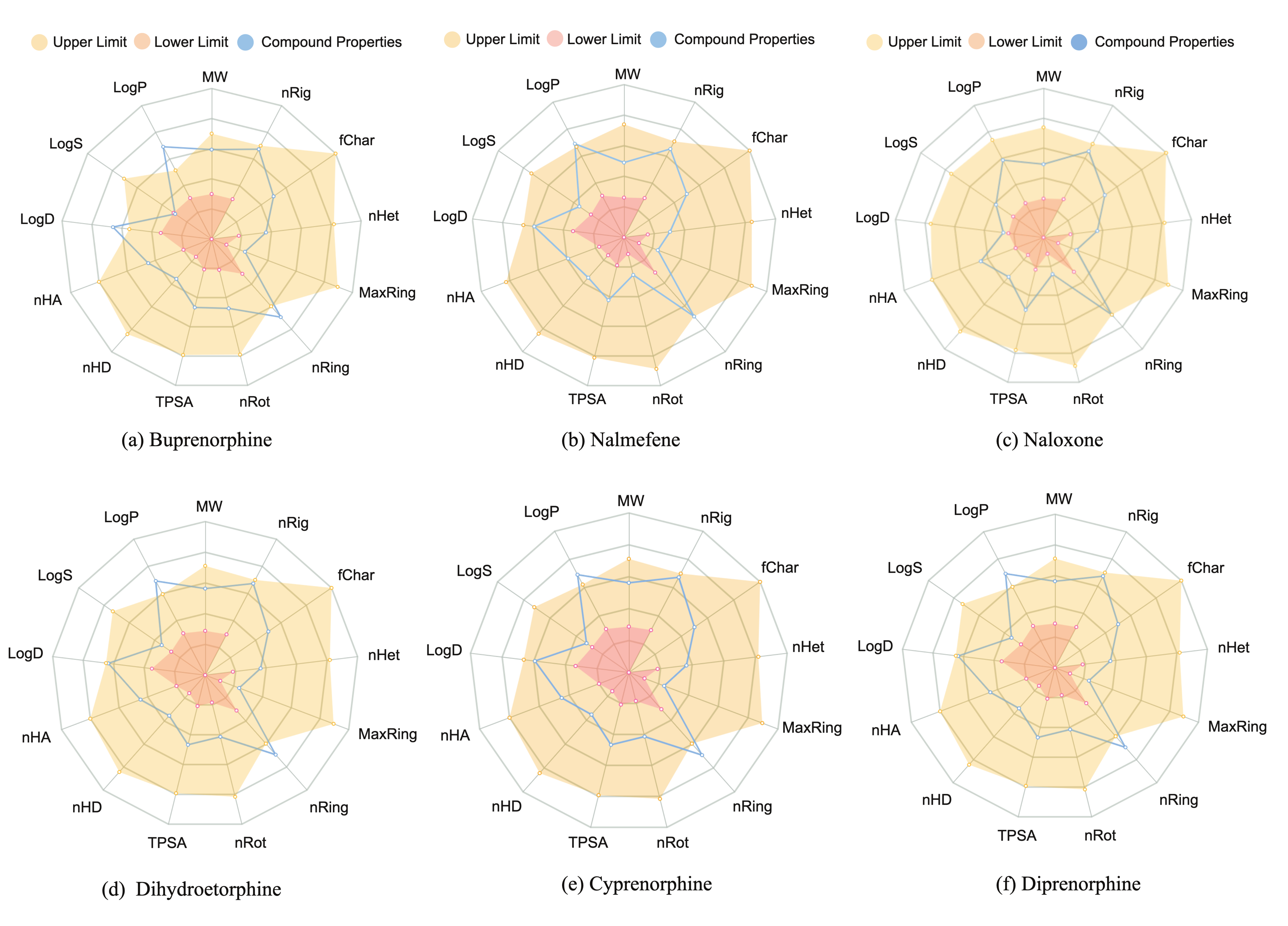} 
			\caption{{\footnotesize Panels a and b show the comparisons of R and RMSE, respectively, for the machine learning models during the five-fold cross-validation tests. The abbreviations used are TF (transformer) and AE (autoencoder). The consensus represents the average of predictions from models built using individual fingerprints. Panel c displays the distribution of experimental and predicted binding affinity (BA) values in the five-fold cross-validation tests of the five machine learning models, with the consensus predictions from the model that uses TF, AE, and ECFP fingerprints being presented. More abbreviations: 
					MW (Molecular Weight), 
					$\log$P (log of octanol/water partition coefficient),  
					$\log$S (log of the aqueous  solubility), 
					$\log$D (logP at physiological pH 7.4), 
					nHA (Number of hydrogen bond acceptors), 
					nHD (Number of hydrogen bond donors), 
					TPSA (Topological polar surface area), 
					nRot (Number of rotatable bonds), 
					nRing (Number of rings), 
					MaxRing (Number of atoms in the biggest ring), 
					nHet (Number of heteroatoms), 
					fChar (Formal charge), and 
					nRig (Number of rigid bonds). The optimal ranges of these indexes  are shown in S3 in the Supporting information.} }
			\label{Fig:ADMET-examples}
		\end{figure}

		The ADMET properties of three FDA-approved and three investigational drugs were predicted using ADMETlab servers, and the predictions are shown in Figure \ref{Fig:ADMET-examples}. Buprenorphine, which has been used to treat OUD, was predicted to have slightly low profiles in the $\log$P (log of octanol/water partition coefficient), $\log$D (logP at physiological pH 7.4), and nRing (number of rings) indexes. The drug's most severe side effect is respiratory depression or decreased breathing. The other two FDA-approved drugs, nalmefene and naloxone, were predicted to be in the satisfactory ranges of all 13 indexes and are useful in counteracting the effects of opioid overdose. The three investigational drugs, dihydroetorphine, cyprenorphine, and diprenorphine, were predicted to have slightly poor values for the $\log$P and nRing indexes. They had similar predictions across all the indexes, which is understandable since these drugs share certain similarities in their molecular structures.
		
		\section{Discussion}

		\subsection{Performance comparisons of molecular fingerprints}
		
		The performance of our ligand-based machine learning models relies heavily on the descriptive abilities of molecular fingerprints. In previous research, we examined the predictive accuracy of traditional 2D fingerprints in forecasting various pharmacological characteristics, such as solubility, partition coefficient, protein-ligand binding affinity, and toxicity \cite{gao20202d}. These fingerprints have proved valuable in drug discovery \cite{gao2020repositioning}. Recently, deep learning models based on natural language processing (NLP) have been created to extract molecular descriptors \cite{chen2021extracting, winter2019learning}, which are also beneficial for machine learning predictions. For this study, we employed one traditional 2D fingerprint and two types of NLP-based molecular fingerprints to build our machine learning models.
		
		We conducted a comparative analysis of the predictive accuracy of three traditional 2D fingerprints and two NLP-based fingerprints for creating machine learning models. We generated the NLP-based fingerprints using transformer (TF) and autoencoder (AE) models, while the traditional 2D fingerprints were ECFP, Estate1, and Estate2. We employed the five inhibitor datasets for opioid receptors and hERG in our modeling comparisons. In addition to modeling with individual fingerprints, we obtained predictions by averaging the predictions from several individual models. These outcomes are referred to as consensus predictions or consensus models, resulting in six models in this study. We constructed three individual models using TF, AE, and ECFP fingerprints, and three consensus models using TF and AE, TF, AE, and ECFP, and ECFP, Estate1, and Estate2. Their performances were compared through five-fold cross-validation. The detailed comparisons can be found in Figure \ref{Fig:fp-comparisons} and Table s2 in the Supporting information.
		
		Of the six models, the consensus model employing TF, AE, and ECFP fingerprints demonstrated the best performance, achieving the highest R values in modeling four out of the five datasets. It exhibited significant improvements compared to models using individual fingerprints or other consensus models. In our previous research \cite{gao20202d}, the consensus model with ECFP, Estate1, and Estate2 displayed high-quality performance in predicting binding affinity. However, the consensus approach with TF, AE, and ECFP proved to be a better option than the consensus model with ECFP, Estate1, and Estate2. Among the three individual models, the ECFP model had higher R values for four out of the five tasks and better RMSE values for three out of the five tasks. Despite the emergence of NLP-based fingerprints \cite{chen2021extracting, winter2019learning}, ECFP fingerprints still demonstrated exceptional predictive ability in modeling the five tasks. They even outperformed the consensus model with TF, AE, and ECFP in terms of R and RMSE values on one and two of the five datasets, respectively. Ultimately, the consensus models with TF, AE, and ECFP fingerprints were chosen to predict the binding affinities of DrugBank compounds on the opioid receptors and hERG since they exhibited the best overall performance.

		\subsection{Virtual screening of hERG side effects}
		
		The hERG potassium channel is a prime target for drugs with diverse structures that can block it and potentially lead to fatal heart irregularities. Hence, evaluating hERG inhibition is crucial in drug development, and machine learning models are excellent tools for virtual screening of hERG side effects. Our machine learning model for hERG inhibition exhibited robust predictive power with R values of 0.786 and RMSE of 0.773 in the five-fold cross-validation. In this study, we used a threshold of -8.18 kcal/mol (K$_i=1\mu$M) to differentiate hERG blockers from non-blockers. Based on our predictions, 408 DrugBank compounds did not pass the hERG screening, with predicted binding affinities (BAs) of less than -8.18 kcal/mol on hERG. The remaining 8457 DrugBank compounds provide a vast pool of drug candidates for repurposing in OUD treatment. Consequently, we predicted their binding energies on the four major opioid receptors.
		
		\subsection{Model reliability associated with label distributions}		
		
		The predictive performance of our ligand-based models relies heavily on the descriptive capabilities of molecular fingerprints. It is essential to have high-quality data with a wide distribution of labels and high diversity of molecular compounds to reduce bias and improve real-world predictions. Our five models exhibited strong predictive power with superior R and RMSE values in the five-fold cross-validation tests, indicating their reliability as binding affinity predictors. Figure S1 in the information section displays the label distributions of the five training sets, which were reasonable except for the NOR dataset. The NOR dataset had low-quality labels with a high percentage of high-affinity values, resulting in a machine-learning model that consistently yields inaccurate predictions in the direction of high-affinity.
		
		In Figure \ref{Fig:fp-comparisons}c, the five-fold cross-validation predictions for the five models are shown, with the predicted BAs from the NOR model having distributions consistent with those of the experimental BAs. However, when estimating the BAs of DrugBank compounds on NOR, our machine-learning model could give overestimated predictions due to the low similarity score of DrugBank compounds with the NOR dataset. The DrugBank compounds had relatively high similarity scores with the MOR, KOR, and DOR inhibitor datasets, but a lower score with the NOR dataset. Therefore, we are only concerned with the binding effect of DrugBank compounds with similarity scores $>$ 0.8 on NOR, and the DrugBank compounds with predicted high BAs are listed in Table S11 in the Supporting information. Unfortunately, addressing the bias in the NOR data or machine-learning prediction is challenging for us.

		\subsection{Dataset element distributions}		
		
Understanding the data distribution is important for featurization. Additionally, the similarity among data distribution is also an important concept in machine learning, as it can impact the performance of various algorithms and techniques. In machine learning, the assumption is often made that the training data is representative of the data that the model will be applied to in the future. This assumption is based on the idea that if the training data is similar to the test data, the model will be able to generalize well to new, unseen data. If the data distribution in the training set is significantly different from the distribution in the test set, the model's performance may suffer. This is because the model will be trained on data that does not represent the types of inputs it will encounter in the future. In such cases, the model may fail to generalize well and may perform poorly on the test set.

			\begin{figure}[ht]
			\centering
			\includegraphics[width=0.95\linewidth]{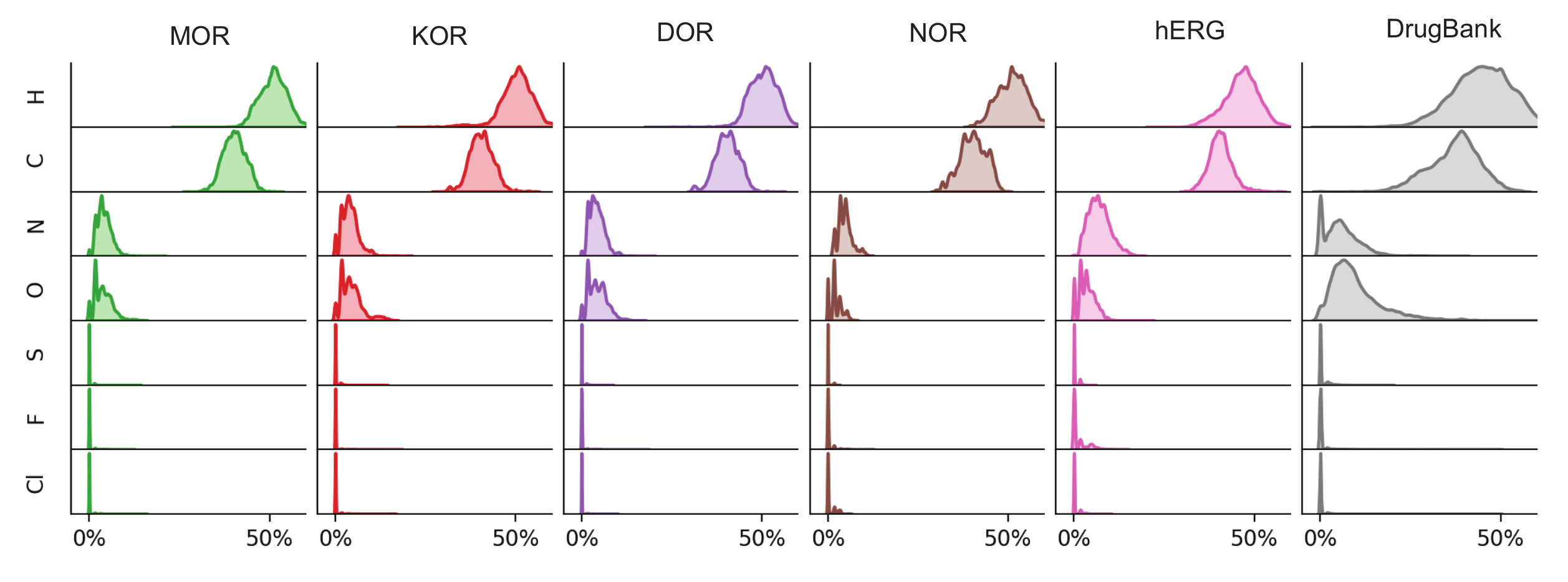} 
			\caption{{\footnotesize The element distributions in molecules of the used six datasets in this study. Elements H, C, N, 0, S, F, Cl are the most dominating elements in all the six datasets.} }
			\label{Fig:stats-elemets}
		\end{figure} 
		
In Figure \ref{Fig:stats-elemets}, we analyzed the element distributions of MOR, KOR, DOR, NOR,  hERG, and DrugBank  datasets. 
The $x$-axis represents the percentage of a specific element in a molecule, while the $y$-axis shows the density of percentages. The elements H, C, N, and O are the most dominant elements observed in these distributions.
We show that MOR, KOR, DOR, NOR, and hERG  datasets  have nearly identical element distributions for their major elements.  These similarities ensure the accuracy and reliability of the our cross-predictions presented in this work. We further analyzed the element distribution of molecules in the DrugBank, as show the last panel of Figure \ref{Fig:stats-elemets}. We noted that the element distribution of DrugBank dataset is also very similar to those of other five datasets, ensuring our machine learning models' ability for repurposing the DrugBank compounds for opioid use disorder.  More information about dataset element distributions is given in Table S1 in the Supporting information.

		\section{Methods}
		
		\begin{figure}[ht]
			\centering
			\includegraphics[width=0.95\linewidth]{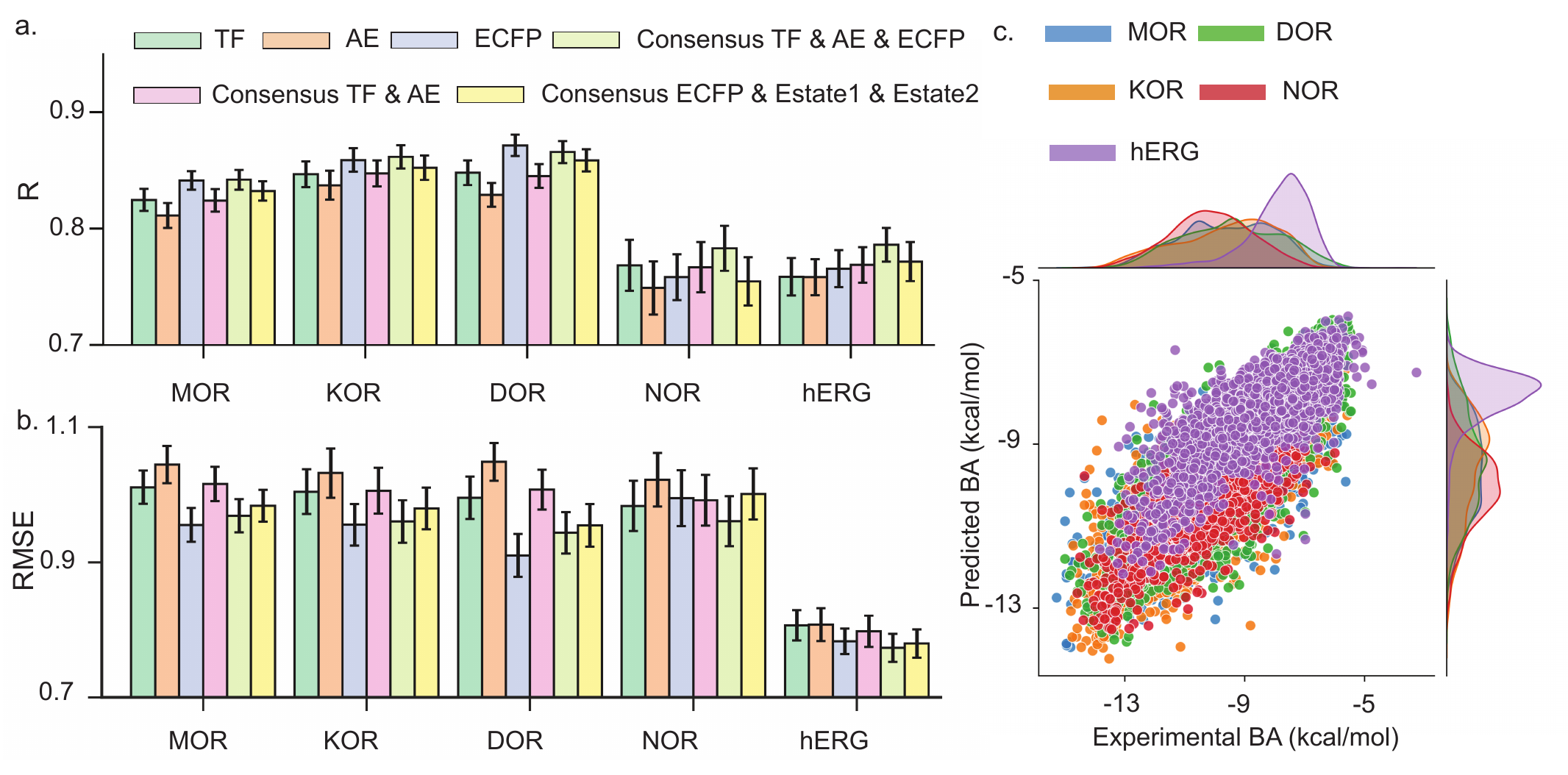} 
			\caption{{\footnotesize Panels a and b present comparisons of R and RMSE values for the machine learning models in the five-fold cross-validation tests. The abbreviations used are TF (transformer) and AE (autoencoder). The consensus represents the average of predictions from models built using individual fingerprints. Panel c displays the distribution of experimental and predicted BA values in the five-fold cross-validation tests of the five machine learning models, with the consensus predictions from the model that uses TF, AE, and ECFP fingerprints.} }
			\label{Fig:fp-comparisons}
		\end{figure}

		\subsection{Data preparation}
		
		\begin{table}
			\centering
			\begin{tabular}{c|c|c|c|c}
				\hline
				\makecell[c]{Dataset} &\makecell[c]{Protein name}   &\makecell[c]{ChEMBL ID }& \makecell[c]{Dataset size} & \makecell[c]{Binding affinity \\range (kcal/mol)} \\
				\hline
				MOR	&Mu opioid receptor& CHEMBL233 & 4667 & [-15.27,-5.60]  \\
				DOR  &Delta opioid receptor& CHEMBL236 & 4033 & [-15.00,-5.46] \\
				KOR&Kappa opioid receptor & CHEMBL237 & 4249 & [-14.80,-5.47] \\
				NOR &Nociceptin opioid receptor & CHEMBL2014 & 1494 & [-14.59,-6.00] \\
				hERG & hERG potassium channel & CHEMBL240 & 6298 & [-13.84,-3.27] \\
				DrugBank & -& -& 8865 & -\\
				\hline
				
			\end{tabular}
			\caption{The summary of all the datasets used in this study. }
			\label{tab:dateset-summary-opioids}
		\end{table}
		
		To build our machine learning models, we obtained the inhibitor datasets for MOR, KOR, DOR, and hERG from the ChEMBL database, which include SMILES strings of molecular compounds paired with corresponding labels. The original labels for the data points were IC${50}$ or $\rm K_{i}$. To convert these experimental labels to binding affinities (BAs) for our models, we used the formula BA=1.3633$\rm \times \log_{10} K_{i}$ (kcal/mol). IC$_{50}$ labels were approximated to K$_i$ values using the relation K$_i$=IC${50}/2$, as recommended by Kalliokoski \cite{kalliokoski2013comparability}. 
		DrugBank   database (version 5.1.10) has  1806 approved drugs and 7059 investigational or off-market drugs, giving rise to a total  8865 compounds  \cite{wishart2018drugbank}.
		A summary of all the datasets used in this study is presented in Table \ref{tab:dateset-summary-opioids}.
		
		\subsection{Molecular fingerprints}
		
		The molecular inhibitors in the five datasets were represented by their 2D SMILES strings, from which three types of molecular features were generated. These features were created using two natural language processing (NLP)-based techniques and a traditional 2D fingerprint. The transformer and sequence-to-sequence autoencoder algorithms were used to generate two types of fingerprints, referred to as TF-FP and AE-FP, respectively. These algorithms utilized pretrained models to generate latent embedding vectors of length 512 from the SMILES strings. Additionally, the traditional 2D fingerprint used was ECFP (extended-connectivity fingerprints), which was generated using the RDKit package \cite{landrum2013rdkit}. The ECFP used in this study had a length of 2048. Further details about the three molecular fingerprints are provided below.

		\subsubsection{Sequence-to-sequence auto-encoder }

		A recent study introduced an unsupervised learning model that can extract molecular information from SMILES representations using data-driven techniques \cite{winter2019learning}. This model uses a sequence-to-sequence autoencoder, which consists of an encoder and a decoder neural network, to compress the molecular information into a latent representation. Physicochemical information can be encoded in the intermediate vectors during the translation process, and the resulting pretrained model can extract molecular descriptors from query SMILES strings without requiring retraining or labels. Additionally, the intermediate latent space vectors between the encoder and decoder can serve as molecular fingerprints for various prediction tasks.
		
		The encoder and decoder networks of the autoencoder model are connected through an information bottleneck that compresses the input SMILES data and feeds the essential information to the decoder. The encoder network uses both convolutional neural network (CNN) and recurrent neural network (RNN) architectures, and fully connected layers map their outputs to intermediate vector representations. The decoder network is primarily composed of RNN networks that receive the intermediate vector representations as input. To incorporate more physicochemical information about molecules in the latent vectors, a classification model was added to predict certain molecular properties based on these vectors. The output of the decoder's RNN network is probability distributions over the different characters in the translated molecular representations. The loss function during training of the autoencoder model is calculated by summing cross-entropies between probability distributions and one-hot encoded correct characters, as well as mean squared errors for molecular property predictions. The model was trained using a large dataset of molecular compounds from the ZINC \cite{irwin2005zinc} and PubChem \cite{kim2016pubchem} databases.

		\subsubsection{Bidirectional transformer }
		
		Chen et al. recently developed a self-supervised learning (SSL) platform to pretrain deep learning models on millions of unlabeled molecules, providing latent space vectors for input SMILES \cite{chen2021extracting}. These vectors contain valuable molecular biochemical information and can be used as molecular fingerprints for predictive tasks. The SSL pretraining was achieved using the bidirectional encoder transformer (BET) model, which relies on the attention mechanism. The BET model is advantageous for parallelism capability and faster training since it avoids the traditional encoder-decoder framework for sequential data processing.
		
		In the SSL pretraining platform for encoding SMILES, pairs of real SMILES and masked SMILES were constructed with a certain percentage of symbols in the strings hidden, and the model was then trained in a supervised way with the data-mask pairs \cite{chen2021extracting}. In order to construct pairs of real and masked SMILES strings for the SSL pretraining platform, 15\% of the symbols in all SMILES were used for data masking. Of these masked symbols, 80\% were fully masked, 10\% were unchanged, and the remaining 10\% were randomly altered. The BET model used in the pretraining platform incorporates an attention mechanism to capture the significance of each symbol in the SMILES strings. To train the model, the Adam optimizer was utilized with a weight decay of 0.1. In this study, the mean of 256 embedding vectors for a given SMILES string was utilized as its molecular fingerprint.
		
		In Chen et al.'s work, SMILES strings from one or the union of the ChEMBL, PubChem, and ZINC databases were used for the SSL-based BET model training \cite{chen2021extracting}. Three models were trained, and a fine-tuning strategy was introduced for these pretrained models for specific downstream predictive tasks. In the current study, the BT-FPs generated directly from the pretrained model on ChEMBL without fine-tuning were used for each dataset.
		
		\subsubsection{Extended-connectivity fingerprints}

		Extended-connectivity fingerprints (ECFPs) are a commonly used type of topological fingerprints for molecular characterization. The ECFP algorithm is based on a variant of the Morgan algorithm that assigns numeric identifiers to each atom through an iterative process. This results in numbering-invariant atom information being encoded into an initial atom identifier, which is then combined with identifiers from the previous iteration to create a unique canonical numbering scheme for the atoms. Unlike the Morgan algorithm, ECFP generation continues until a predetermined number of iterations is reached. The initial atom identifiers and all intermediate identifiers in each iteration are collected into a set, which defines the extended-connectivity fingerprint \cite{rogers2010extended}. We used the RDKit library \cite{landrum2013rdkit} to generate ECFPs, which builds circular fingerprints based on the Morgan algorithm and requires a radius parameter that determines the number of iterations the algorithm should perform. In our implementation, we set the ECFP radius to 3 and the length of the ECFP fingerprint to 2048.
		
		\subsection{Machine-learning models}

		Our machine learning models were developed using the gradient boosting decision tree (GBDT) algorithm, which is known for its robustness against overfitting, insensitivity to hyperparameters, and ease of implementation. The GBDT algorithm creates multiple weak learners or individual trees by bootstrapping training samples and integrates their outputs to make predictions. Although weak learners are prone to making poor predictions, the ensemble approach can reduce overall errors by combining the predictions of all the weaker learners. GBDT is particularly useful for training with small datasets and has been observed to outperform deep neural networks (DNNs) and other machine learning algorithms in a range of quantitative structure-activity relationship (QSAR) prediction problems \cite{feng2023virtual,pun2022persistent}, making it a popular choice for developing predictive ML models. We used the GBDT algorithm provided in the Scikit-learn (version 0.24.1) library for our work.
		
		The study employed three types of molecular fingerprints (TF-FP, AE-FP, and ECFP) to represent inhibitor compounds and built three separate machine learning (ML) models using the gradient boosting decision tree (GBDT) algorithm. To enhance the reliability and robustness of the models for predicting binding affinities (BAs) for MOR, KOR, DOR, NOR, and hERG, a consensus strategy was employed. The consensus model was constructed by taking the average of the predicted BAs from the three individual models. This approach is commonly used to improve predictive performance and has been shown to be more effective than using individual models alone \cite{zhang2022hergspred,gao20202d}. Figure \ref{Fig:fp-comparisons} presents a comparison of models using various fingerprints or modeling strategies based on five-fold cross-validation tests. The results indicate that the consensus models outperformed the individual models, and the consensus model using all three fingerprints showed the best predictive ability for the five datasets. To alleviate the effects of randomness, each individual model was built ten times using different random seeds. In total, the 30 predictions from the three individual models were used to calculate the consensus BAs, which were used as the predicted binding affinities for each protein.
		
		\section{Conclusion}
		
		Opioid use disorder (OUD) is a chronic and complex disease characterized by compulsive and uncontrollable opioid use, leading to physical and psychosocial disruptions. In response to the worsening substance use crisis in the United States, significant efforts are being invested to combat this issue, especially during the coronavirus pandemic. However, the effectiveness of currently available medications for OUD is limited by low utilization rates and high relapse rates. Hence, there is a pressing need for medications with superior therapeutic effects that can prevent relapse and promote longer periods of abstinence. Repurposing existing drugs could expedite the development of additional medications.
		
		Machine learning-based computational approaches can systematically and rapidly screen the repurposing potential of many drugs at a low cost. Opioid receptors, including MOR, KOR, and DOR, are the primary pharmacological targets of medications used to treat OUD. In this study, we curated inhibitor data from the ChEMBL database to build machine-learning predictive models of binding affinity. Using these models, we carried out predictions and analyses to identify DrugBank compounds that can selectively bind to different opioid receptors. We discriminated approved or investigational drugs based on various binding affinity thresholds to identify repurposable candidates. We focused on drugs that have demonstrated pharmacological effects in treating OUD and further analyzed their molecular interactions by docking with the receptors. We screened ADMET properties using machine learning-based models. However, a candidate's therapeutic efficacy in OUD treatment needs to be further investigated for many other indexes, including antagonist/agonist effect and blood-brain barrier permeability, which partially explains the complexity of drug design for the treatment of substance use disorder.

		Machine learning has shown potential as a valuable tool in aiding drug discovery efforts for the treatment of OUD. One approach involves leveraging approved drugs or other DrugBank compounds with repurposing potential as a starting point to design new candidate compounds, which can be assisted by generative network modules \cite{gao2020generative}. Additionally, machine learning-based virtual screening approaches are increasingly being used for opioid drug discovery \cite{jia2021construction, sakamuru2021predictive}. These machine learning studies have the potential to offer valuable insights and assist in the development of pharmacological treatments for OUD.

		\section*{Data and code availability}

	The related datasets studied in this work are available at: 
	https://weilab.math.msu.edu/DataLibrary/2D/. Codes are available at https://github.com/WeilabMSU/opioid-repurposing. 

		\section*{Supporting Information}
	
		The Supporting information includes
		 S1 Datasets and model performance summary, 
		 S2 Drug similarity analysis, S3 ADMET indexes and the optimal ranges, and S4 Additional prediction results for DrugBank compounds.

		\section*{Acknowledgment}	
	This work was supported in part by NIH grants R01GM126189 and R01AI164266, NSF grants
	DMS-2052983, DMS-1761320, and IIS-1900473, NASA grant 80NSSC21M0023, MSU Foundation, Bristol-Myers Squibb 65109, and Pfizer.

		%
		%
		\section*{References}

\end{document}